 \documentclass[twocolumn]{article}
\usepackage{graphicx}

\begin{document}

\title{Arctic East Siberia had a lower latitude in the Pleistocene} 

\author{W. Woelfli\footnote{ Institute for Particle Physics, ETHZ H\"onggerberg, CH-8093 Z\"urich, Switzerland (Prof.~emerit.); e-mail: woelfli@phys.ethz.ch .} \ and
W. Baltensperger\footnote{Centro Brasileiro de Pesquisas F\'\i sicas, Rua Dr.\thinspace Xavier Sigaud,150, 222\thinspace 90 Rio de Janeiro, Brazil;\qquad \qquad e-mail: baltens@cbpf.br}}
\date{March 30, 2006}
\maketitle

\noindent\emph{In Arctic East Siberia many remains of mammoths have been found. In this region there is not sufficient sunlight over the year to allow for the growth of the plants on which these animals feed. Consequently the latitude of these regions must have been lower before the end of the Pleistocene than at present. It is a challenge to reconstruct this geographic shift of the poles in a manner compatible with known facts. A possible sequence of events is described here. It assumes an additional planet, which must since have disappeared. This is possible, if it moved in an extremely eccentric orbit and was hot as a result of tidal work and solar radiation. During a few million years evaporation of this planet led to a disk-shaped cloud of ions moving around the Sun. This cloud partially shielded the Earth from the solar radiation, producing the alteration of cold and warm periods characterizing the Pleistocene. The degree of shielding is sensitive to the inclination of Earth's orbit, which has a period of 100\thinspace 000 years. Two  cloud structures are discussed. The first is small and steady. The other builds up to a point where inelastic collisions between  particles induce its collapse The resulting near-periodic time dependence of the shielding resembles that of  Dansgaard-Oeschger events.  The Pleistocene came to an end when the additional planet had a close encounter with the Earth, whereby the Earth suffered a one permil  extensional deformation. While this deformation relaxed to an equilibrium shape in a time of one to several years, the globe turned relative to the rotation axis: The North Pole moved from Greenland to the Arctic Sea. The additional planet split into fragments, which subsequently evaporated. Simple estimates are used here for the characterization of the complex processes; more elaborate studies may lead to different scenarios for the indispensable pole shift.}

\section{Reveiling evidence}
The remains of mammoths in East Siberia in regions with high latitude have been found in ecological surroundings, where at present these herbivores could not exist. In East Siberia heards of mammoths grazed within the arctic circle, even on islands in the Arctic Sea which during the glacial periods were connected with the mainland. The yearly insolation diminishes with increasing latitude, and the distribution of the flora on the globe suggests that in arctic regions  the yearly insolation is insufficient for the steppe plants that feed mammoths. A limiting situation exists in the Wrangle Islands, where in some favourable habitats certain depauperated relicts of the Pleistocene grassland supported the persistence of dwarf mammoths into the Holocene \cite{Vartanian}. Floral reconstructions for the Late Pleistocene have also been made on the basis of fossil beetles in arctic East Beringa \cite{Alfimov}. This paper is written with the assumption that beyond a limit given essentially by the arctic circle the insufficient yearly insolation inhibits the growth of the steppe plants indispensable for mammoths. Why then could mammoths exist in arctic areas? Since this depends on the yearly insolation,  there is just one answer: these regions had a lower latitude in the Pleistocene. 

Let us suppose that the North pole was at the center of the known ice cover of the Last Glaciation. This is situated in Greenland, about $18^\circ$ apart from the present North pole. Its longitude is less certain, since it depends on the thermal influence attributed to the Atlantic ocean. A model study of the Pleistocene climate with geographically shifted poles could be revealing. The geographic consequences of a polar shift are best visualized using a globe.  The angular distance from the old or new pole amounts to (90$^\circ$ - latitude) in each case. Since the North pole moved from Greenland into the Arctic Sea while the South Pole was  displaced within Antarctica, the climate changes were larger on the Northern hemisphere. Some places on the great circle through the old and the new positions of the poles suffered the full 18$^\circ$ shift. The Lena River in Siberia moved  18$^\circ$ north, while the latitudes in Australia decreased approximately by this amount. Bolivia moved away from the equator (tropical $\rightarrow$ arid), while the Northern Amazon region shifted to the equator (arid $\rightarrow$ tropical). The latitudes on the US-East coast and in West Europe were higher in the Pleistocene, and those of Alasca slightly lower. 

The evidence of mammoths in arctic east Siberia is not just one more in a multitude of unexplained facts, since it contains an aspect that we understand. These regions necessarily received more sunlight in the Pleistocene than at present.  Thus, the latitude of arctic East Siberia was lower than it is now. The globe has been turned with respect to its rotation axis.

If the idea of the geographic shift of the poles appears exceptional, then so are the empirical facts. (They are of a kind similar to the data about photoemission as known at the beginning of the last century.  The Maxwell equations then continued to be valid, however,  something additional [i.e. quantization] was necessary for the explanation.) At present, the Milankovitch theory continues to be valid, but without a lower latitude of East Siberia during the Pleistocene the known facts cannot be understood. The evidence imposes a conclusion which adds a basic assumption i.e. a shift of latitude of Arctic East Siberia. Since this geographic pole shift actually took place there must be at least one possible scenario to produce it. In the following paragraphs we attempt to find such a course of events. This involves quite complex processes for which we can only provide simple estimates. If more elaborate future studies will show that our estimates are wrong,  this means that the real scenario must have been different, but not that the shift of the poles did not occur.

The problem discussed here has a long history. At the end of the 19$^{th}$ century, several geographers postulated a polar shift on the basis of the asymmetry of the observed Pleistocene maximum glaciation. Careful studies by G. H. Darwin, J.C. Maxwell,  G.V. Schiaparelli as well as by  W. Thomson led to the conclusion that the required rapid polar shift was impossible. The verdict of these eminent scientists was considered definite. At that time, condensed matter was considered to be either solid or liquid. During the last century the concept of plastic behaviour appeared which opened a new range of possible relaxation times of deformation. It may have remained unnoticed that this is of decisive importance for the problem of a rapid geographic polar shift. Without the polar shift, the problem of mammoths in polar regions remains unsolvable.

\section{Geographic polar shift}

A rapid geographical shift of the poles is physically possible \cite{Gold}. At present, the Earth is in hydrostatic equilibrium. Since it rotates, its radius is larger at the equator (by 21 km) than at the poles. The rotational motion of an object is governed by its inertial tensor. In a coordinate system fixed to the object and with the origin at the center of mass, this tensor is obtained by an integration over the density times a bilinear expression of the cartesian components. At present, due to the equatorial bulge, one of the main axes of Earth's inertial tensor is longer than the other two, and its direction coincides with that of the rotation axis. This is a stable situation. For a polar shift, a further deformation of the Earth is required. During the shift, the direction of the angular momentum vector remains strictly fixed relative to the stars, as required by conservation laws. What turns is the globe relative to the rotation axis. The shift leads to new geographic positions of the North and South Poles. 

Suppose the Earth gets deformed: some of its mass is displaced to an oblique direction. This produces an inertial tensor with a new main axis, which deviates  from the rotation axis. Then, as seen from the globe (i.e. geographically), the rotation axis will move around ("precess" around) this main axis. Actually, on a minute scale such a precession is observed on the present Earth (Chandler precession). A full turn of the precession takes about 400 days. This period is determined by the equatorial bulge. Its order of magnitude will be of prime importance in the discussion of polar shifts. The shape of a deformed Earth relaxes to a new hydrostatic equilibrium, which brings the precession to an end. In the final situation there again is an equatorial bulge around the new geographic position of the rotation axis. The poles have shifted geographically, however, what turned in space is the globe. 

If global deformations relax in  a time short compared to about 200 days (half a precession cycle), the movement stops quickly and the pole shift remains insignificant. This would happen in the case of an elastic deformation of a solid Earth, since the changes of the deformation occur with the speed of sound. Similarly, for a model of a liquid Earth, the backflow of matter (over distances of 10 km) is expected to occur within days at most. Historically, at the end of the 19$^{th}$ century, these were the only known states of condensed matter. Thus, a polar shift seemed to be impossible \cite{Hapgood}. In the last century, plastic materials with wide ranges of relaxation times were investigated. The idea that a global deformation of the Earth relaxes in several years appears plausible. A simple calculation  of a geographic polar shift with just one assumed relaxation time of 1000 days is given in the Appendix of ref.\cite{planet}. The result is a decreasing precession of the rotation axis around the main axis, which itself moves on a small spiral. Of course, a study of the motion with a more detailed model of the Earth would be very significant.

\vspace{0.5cm}\section{Cause of the deformation}

The polar shift requires a displacement of mass on Earth at the end of the Pleistocene. What mechanism could produce this? Hapgood \cite{Hapgood} proposed that the ice on Antarctica could become unstable and drift away from the South Pole due to the centrifugal force. For an appreciable shift, the ice would have to move several tens of degrees latitude and add to the mass of a continent rather than float. Even then, this displacement of mass could actually not produce the required shift. A related idea might consider a displacement of the Earth's nucleus from the centre by centrifugal forces. Again, this displacement would have to be large, and it is incompatible with the present centred position of the nucleus.  

A very efficient deformation is a stretching of the globe in a direction oblique to the poles. A volume flow over distances of the order of the streching amplitude suffices to create a mass difference at the surface far from the rotation axis. For the pole shift considered, the required stretching amplitude is  $ 6.5$ km on each side \cite{planet}. How could this one per mil stretching in a direction $30^\circ$ from the poles occur? If a massive object passed near the Earth, it would create a tidal force. Since (for large distances) tidal forces vary with the third reciprocal power of the distance to Earth, the  Moon brought 20 times closer would produce a large but still insufficient tidal effect. A close passage of a mass about ten times larger is required (approximately the mass of Mars) \cite{Notas1}. It is reasonable to assume a planetary speed for this object, say a relative velocity to the Earth of about 40 km/s. The close distance then lasts about 10 minutes only. The process of deformation is therefore highly dynamic. Only an elaborate study of this process could give reliable  numbers. The required one per mil value is the deformation with a prolonged relaxation of at least a hundred days. Any  deformation that decays rapidly would be additional. This global deformation is catastrophic, although compatible with the continuation of life on Earth. Nevertheless, many large vertebrate species are known to have become extinguished at the end of the Pleistocene \cite{Martin}.  

\section{Planet Z}

What was this massive object, that passed near the Earth? Certainly not one of the present planets, since these have orbits that do not pass through Earth's distance from the Sun. The object involved in the near collision must afterwards have been in an orbit, which crosses that of Earth, and the Holocene was much too short a time for a major readjustment due to couplings with Jupiter and other planets. One might think of a massive object which happend to travel through the planetary system. This might occur as a rare event. However, it is improbable that at this occasion the object comes close to Earth. For this reason the object has to be an additional planet in an orbit that  crosses Earth's distance from the Sun. The chance that a passage through the surface of the sphere at Earth's distance, $R_E=150$ Mio km, happens within a range of $r=20\thinspace 000$ km  is only $\pi r^2/(4\pi R_E^2)=4\cdot 10^{-9}$. Thus it is even unlikely that within the time in which an object may remain in an exotic orbit, i.e. a few million years, a narrow encounter occurs. Therefore we assume that the orbital plane of this planet, henceforth called Z, is restricted to a small angle (say $ 1^\circ$) with the invariant plane (perpendicular to the total angular momentum of the planetary system). For example Z might have been a moon of Jupiter which got loose. 

The larger the distance of an approach between Z and Earth, the  more frequent it is.  A passage of Z nearer than about the distance Earth-Moon creates dramatic  earthquakes.     These may well  be the trigger of Heinrich events \cite{Heinrich}.

The orbital parameters of Z are not known, but restricted by three conditions. The perihel distance leads to a hot planet. The aphel lies beyond Earth's orbit. Its value together with the variable inclination of Z to the ecliptic allow a close approach to Earth within some million years. In numerical estimates we often used $4\cdot 10^9$ m for the perihel distance and $1.5\cdot 10^{11} $ m for the semi-axis of the ellipse. This corresponds to an eccentricity $\epsilon = 0.973$.

\section{Disappearance of Z}

Evidently, at present Z does not exist. How could it disappear within the Holocene?  Only the Sun could accomplish this. Z had to be in a special situation before the pole shift, i.e. during the Pleistocene. Necessarily, Z had to move in an extremely eccentric orbit, with a perihel distance barely compatibel with its existence. Each time Z passed through the perihel, it was heated inside by tidal deformation and on the surface by solar radiation. Z was liquid and had a shining surface.

Since Z must disappear during the Holocene, it is almost indispensible that Z broke into pieces during the narrow passage. For this Z must be much smaller than Earth, so that the tidal forces produced by Earth on Z are  larger than those by Z on Earth. The  pressure  release in the hot interior may have  further promoted the preakup. The condition that Z had at least 1/10 of the mass of Earth, so that it could deform the Earth as required by the polar shift, together with the condition that it was much lighter than Earth, imposed by the breakup, determine the size of Z surprisingly well. Z was about Mars-sized. Again, these considerations deserve detailed studies.
  
For an evaporation from Z,  the particles have to surmount their escape energy. The escape speed of Mars is 5.02 km/s. As an example, the kinetic energy of an Oxygen atom (as the most frequent atom on a dense planet) with that speed is 2.1 eV and that of an O$_2$ molecule 4.2 eV. If half the molecular binding energy of an O$_2$ molecule is included as the cost to produce an evaporated O-atom, it turns out that it takes less energy (4.2 eV for O$_2$ versus 4.7 eV for O) to evaporate the molecule than the single atom. However, note that this holds, since O$_2$ is a fairly light molecule. In most cases, atomic evaporation prevails. In a theory of evaporation, the Boltzmann factor $\exp  [-E/(k_BT)]$ plays a dominant role, where $E$ is the energy necessary to liberate a particle, $T$ the temperature and $k_B$ the Boltzmann constant. Let us assume $T=1500$ K on the surface of Z near the perihel. For $E$ we use the escape energy for an Oxygen molecule from Z, i.e. $E_1 = 4.2$ eV. If Z breaks into $n$ equal parts Z$_n$, each has $1/n$ the mass of Z, while its radius is reduced by $(1/n)^{1/3}$ at most.  Therefore the  escape energy  from a fragment satisfies $E_n  \leq E_1/n^{2/3}$, and the ratio between the Boltzmann factors for Z$_n$ and  Z becomes 
\begin{equation}
e^{{E_1-E_n\over k_BT}} \geq \left\{ \begin{array}{ll}2\cdot 10^5 &\mbox{ for } n=2\\2\cdot 10^7 &\mbox{ for } n=3\end{array}\right. 
\end{equation}
Thus the splitting of Z into two or more parts results in an enormous increase of the Boltzmann factor. Mostly for this reason, we expect a dramatic increase of the evaporation rate after the polar shift. From the fractions Z$_n$, molecules and clusters evaporate. Furthermore, since the near-collision  between Z and Earth dissipates energy, it is likely that the  perihel distances of the parts Z$_n$ are reduced. If their masses diminish sizeably in the following 1000 years, then the complete evaporation within the Holocene results. Obviously, these considerations are preliminary. They indicate the possibility that Z can disappear within the Holocene, provided that it was already evaporating during the Pleistocene. 

In principle, Z could vanish in a different way. Since it was in an extremely eccentric orbit, there is a certain probability that during the narrow encounter it lost its small angular momentum and afterwards dropped into the Sun. However, as Bill Napier pointed out in a private communication,  as a result of the attraction to the Sun, a Mars-sized object would introduce a kinetic energy equivalent to the  solar radiation of three years, and furthermore, the shock might perturb the delicate equilibrium in the Sun's innermost parts and activate an increase of the nuclear reaction. For these reasons, we favour processes which gradually add the material of Z to the Sun  over many years.

This assumed scenario for the Pleistocene could not have occured several times during the existence of the planetary system without a collision of Z with one of the inner planets. The Pleistocene ice age era was a rare, if not unique, period in Earth's history.  Other types of ice ages may have occured  as a result of the slow movements of the continents. When these were joined to one block, Earth's rotation is stable when this supercontinent is centered around a pole. Plausibly, in this situation the whole continent is covered by ice. 

\section{The steady gas cloud} 

The continuous evaporation from Z produced a gas cloud.  Since the velocity of Z at the perihel is much larger than thermal particle velocities, the initial conditions of the particle motions equal those of Z. However, each type of particle is subject to its specific light force. If the first excitation energy of the particle is larger than about 10 eV, the repulsive light force is expected to be weaker than the gravitational attraction to the Sun. In this case bound orbits exist. In general, an evaporated particle may start on a hyperbolic or an elliptical path. 

The properties of the cloud determine the most important consequences of this model, since the cloud can partially shield  the solar radiation from  Earth. This becomes the prime reason for the glaciations  during the Pleistocene. The cloud is very complex. It involves particles,  each with its light force, plasma properties and possibly magnetic and electric fields. The cloud receives particles that evaporated from Z, which itself is in a time-dependent orbit.  Both, the orbital and the spin periods of Z appear in the evaporation. Furthermore, the dynamics of the cloud itself may produce  time dependencies. The authors have repeatedly changed their views regarding the spacial extent of the cloud and the amount of material lost to outer space. 

In the ice cores of Greenland \cite{Mayewski} and of Antarctica \cite{EPICA} the impurity concentration is sharply peaked during cold periods. Given the uncertainty about the size of the cloud, this dust might have contained extraterrestrial contributions, since during cold periods the Earth might have been in the cloud. However, the detailed analysis  \cite{Biscaye,Delmonte} unambiguously lead to  the terrestrial origin of the dust. Hence, the cloud does not reach Earth$^\prime$s distance from the Sun. It has been a main motivation of this paper to examine whether the present model is compatible with  this empirical fact. 

Since the light force is directed outwards, an individual particle orbit can only extend further out than the orbit of Z. If, empirically, particles do not reach Earth´s distance from the Sun, then they must have lost energy in the first part of their orbit. This could be due to inelastic collisions between particles. Photon emission reduces the relative kinetic energy of two colliding particles. After the collision, the difference of their velocities is smaller, and their new orbits differ less.  However, the sums of the two momenta and also of the two angular momenta remain practically unchanged. When electrons are emitted during the collision, this also holds to a good approximation. The angular momentum of the evaporated particles will define a size of a cloud, which cannot be further reduced by inelastic collisions. 

The solar radiation acts on a particle by a repulsive radial force $F_{rad}$, which varies with the inverse square of the distance to the Sun, just as the gravitational attraction to the Sun, $F_G={mM_S G\over r^2}$, where  $M_S$ is the mass of the Sun, $m$ the mass of the particle and $G$ the gravitational constant. Let,  for a given particle, 
\begin{equation}
f=\left| {F_{rad}\over F_G}\right|
\end{equation}
be the ratio between these forces. Then $G (1-f)$ is a coupling constant, which describes the diminished attraction to the Sun. 
The energy $E$ of a bound particle  depends only on the semiaxis $a$ of its orbit  
\begin{equation}
E=-{G (1-f) m M_S\over 2a}.\label{E}
\end{equation}
 The angular momentum $L$ depends also on the eccentricity $\epsilon$:
\begin{equation}
L=m\sqrt{M_SG (1-f) (1-\epsilon^2)a}.\label{L}
\end{equation}
When the energy $E$ decreases, $a$ decreases, but since $L$ remains constant on the average, so does $\sqrt{(1-\epsilon^2)a}$. Therefore the eccentricity diminishes from its initial value $\epsilon$. This may go on until the eccentricity vanishes and $a$ reaches the value $a_{final}$
\begin{equation}
a_{final} =(1-\epsilon^2)a.
\end{equation}
For $1-\epsilon \ll 1$, this is twice the initial perihel distance $(1-\epsilon )a$. 
Therefore, as a result of assumed collisions in a region quite close to the Sun, particles can end up in circular orbits that differ strongly from the initial orbits of the evaporated particles. With an assumed perihel distance for Z of 4 Mio.~km and a radius of the Sun of 0.7 Mio.~km, $a_{final}$ is only 11 solar radii and 1/11 of the perihel distance of Mercury. If it turns out that the mean free path for collisions is small compared to the size of this cloud, then the assumption that collisions inhibit the escape of particles to Earth's distance becomes self-consistent. 

The kinetic energy of a particle of mass $m$ and evaporated with the speed of Z at perihel is
\begin{equation}
E_{kinetic} = {mM_SG\over 2a}\cdot {1+\epsilon\over 1-\epsilon},
\end{equation}
where $a$ and $\epsilon$ refer to Z. 
The first factor, assuming $a=1.5\cdot 10^{11}$ m, i.e. Earth's orbital radius, corresponds to a velocity of $3.2\cdot 10^4$ m/s and to a  kinetic energy for an Oxygen atom of  85 eV.  With $\epsilon = 0.97$, the second factor multiplies this energy by 66. Therefore,  evaporated particles and also particles in the cloud have enough energy to split molecules and ionize atoms in inelastic collisions. For the resulting ions, the first excitation energy is usually sufficiently large, so that the light force is small compared to the gravitational attraction to the Sun. Collisions can produce ions in circular orbits bound to the Sun.

\section{Lifetime of the particles}

The subsequent lifetime of these particles is determined by  Poynting-Robertson drag \cite{Gustafson}. A particle moving in a circular orbit with velocity $v$  is acted upon not only by the radial lightforce $F_{rad}$  but, due to the aberration, also by a drag force ${v\over c}F_{rad}$ (in linear approximation in ${v\over c}$).  The energy loss per unit time of the particle becomes ${dE\over dt}=-{v^2\over c} F_{rad}$.   Then, using the radius $r$ as time dependent parameter given by $ E = -{mM_SG(1-f)\over 2r}$,
\begin{eqnarray}
{dr\over dt} &=& {dr\over dE}{dE\over dt} \\
&=& -{mM_SG(1-f)\over 2E^2}{v^2\over c}{mM_SG\over r^2}f\\
&=&-{v^2\over c}\cdot {2f\over 1-f}.
\end{eqnarray}
With $v^2=M_SG(1-f)/r$,
\begin{equation}
dt =-{cr dr\over 2fM_SG},
\end{equation}
which integrates to the lifetime $t_0$ for the change of $r$ from $r_1$ to $r_2$
\begin{equation}
t_0 = {c(r^2_{1}-r^2_{2})\over 4M_SGf}.\label{lifetime}
\end{equation}
Here, $r_{2}=0.7$ Mio km, the radius of the Sun, is numerically almost insignificant. With $r_{1}= 1\mbox{  AU } =1.5\cdot 10^{11}$ m , equation [\ref{lifetime}] would give $t_0 ={406\over f}  \mbox{ yr}$, while the distance after the collisions, $r_{1} =a_{final}= 8$ Mio  km, results in $t_0 = {1.1\over f} \mbox{ yr }$. For ions often ${1\over  f} \gg 1$.  For a given ionic state, the value of $f$ is not readily available. Furthermore, it may be difficult to predict the distribution of ionic states that result from the collisions and from  the surrounding illuminated plasma.

The lifetime $t_0$ corresponds to an idealized model, in which the evaporated particles first rapidly loose energy by collisions until they reach circular orbits. Their relative velocity is then assumed to be small, so that inelastic collisions become rare and Poynting-Robertson drag determines the lifetime. In reality, inelastic collisions still occur with newly evaporated particles and possibly also between particles in circular orbits with different values  of $f$. Therefore, $t_0$ is an upper limit only. 

\section{Shielding of the solar radiation}

The amount of solar radiation scattered per unit time by a particle is proportional to the value $f$ of this particle. On the other hand, its Poynting-Robertson lifetime $t_0$ is inversely proportional to $f$. Therefore, the amount of light scattered by a particle during its lifetime is independent of $f$. The more it scatters, the shorter it lives. This allows to give an estimate of the number of particles that have to enter the cloud per unit time to produce a certain  amount of shielding.

The fact  that in equation [\ref{lifetime}] the radii are squared in the numerator  shows that the particles spiralling in the plane of the orbit spend equal time per equal surface. According to equation [\ref{lifetime}] this time per unit surface is
\begin{equation}
\beta ={c\over 4\pi M_SGf}.
\end{equation}
Numerically, $\beta ={1\over f}\cdot 1.8\cdot 10^{-13}$ s m$^{-2}$.

The Sun produces $S= 3.8\cdot10^{26}$ Watt of radiation output. The momentum flow (i.e. the pressure) at distance $r$ is $S/(4\pi r^2c)$. The value $f$ corresponds to a scattering cross section $\sigma$ (assumed as isotropic).
\begin{equation}
\sigma {S\over {4\pi r^2c}}= f\cdot{mM_SG\over r^2}.
\end{equation}
Thus
\begin{equation}
\sigma =f\cdot {4\pi mc M_SG\over S}.
\end{equation}
For an Oxygen  atom (or ion) with mass $m= 2.7\cdot 10^{-26}$ kg this amounts to $\sigma = f\cdot 3.5\cdot 10^{-23}\mbox{ m}^2$. 
 
 A particle spends a time $\beta 2 \pi r dr$  between the distances $r+dr$ and $r$. During this time an individual particle scatters the fraction $\sigma /(4\pi r^2)$ of the outgoing solar radiation. The actual amount of shielding depends on the shape of the cloud. An isotropic cloud around the Sun cannot screen any radiation from an object outside. If immersed into such a cloud, an object would even be exposed to additional radiation from backscattering. 

It is therefore essential to investigate the shape of the cloud, which  depends on the supply of particles, the distribution of $f$ values, the temperature,  the presence of magnetic and electric fields, and the influence of the planets. This is a very complex problem. An inelastic collision reduces the relative velocity between the two colliding particles, so that their orbits become more alike. We assume that the cloud is a disk with the Sun at its center, extending to an outer radius $R$.  Outwards its width increases conically, being limited by a small angle $\alpha$ on both sides of its central plane. The orientation of this plane may be determined by the orbit of Z or vary under the influence of the planets. Possibly it coincides with the invariant plane \cite{Muller1,Muller2} of the planetary system, since in this case the climate varies with the inclination of Earth's orbital plane, which has a period of approximately 100 kyr. Even for the simple assumed shape of the cloud, the shielding of an extended Sun presents complications. In the face of all the uncertainties, we shall simply assume an angle-independent shielding for directions within the cloud and no shielding outside.

Let $\nu$ be the rate by which particles are introduced into the cloud. A particle at distance $r$ from the Sun scatters the fraction  $\sigma/[4\pi r^2 \sin(\alpha )]$ of the light passing through the cloud. In the Pleistocene the temperature differed a few percent between stadials and interstadials, so let us assume a small  shielding $s$, say $s =3\ \%$. Any overlap of crossections can then be neglected. The shielding $s$ is an integral over the contributions of the scatterers at all distances and is assumed independent of direction  within the cloud:
\begin{equation}
s=\int_{r_{2}}^{r_{1}} {\nu\beta 2\pi r dr\sigma\over 4\pi r^2\sin(\alpha )}  ={\nu \beta\sigma\over 2\sin (\alpha )}  \ln \left( {r_{1}\over r_{2}}\right),
\end{equation}
thus
\begin{equation}
\nu = {2\sin (\alpha )\over \ln \left( {r_{1}/ r_{2}}\right)}  \cdot {sS\over mc^2}.\label{nu}
\end{equation}
($\nu mc^2$ is the energy of matter introduced into the cloud per unit time and $\sin (\alpha ) sS$ the total  scattered radiation energy per unit time. Equation [\ref{nu}] states that these quantities differ only by a factor of order 1, when the lifetime of the particles is determined by Poynting-Robertson drag.)  As a numerical example, let's put  $\alpha = 1^\circ={\pi \over180}$  rad , $r_1 = a_{final} = 8\cdot 10^9$ m , $r_{2}= 7\cdot 10^8$ m the radius of the Sun, further $s=0.03$, and finally $m= 2.7\cdot 10^{-26}$ kg, the mass of an Oxygen ion. We then obtain $\nu =7\cdot 10^{31}$ particles/s $= 1\cdot 10^8$ mol/s. For Oxygen this corresponds to a time  averaged mass evaporation rate of $2 \cdot 10^6 $ kg/s. If Z has a radius of $3.4\cdot 10^6$ m and a surface density of 3000 kg/m$^2$, then  a surface layer of 0.1 mm per year evaporates. In 3 Mio years this amounts to 400 m. With this estimate, the size of Z is hardly affected by the evaporation required to diminish the solar radiation within the solid angle of the cloud by 3 \%. 
 
There are several corrections to this evaporation rate. Due to the finite size of the Sun, the rays that move towards  the shielded Earth mostly enter the cloud at a distance larger than the Sun's radius. Thus  the 3 \% shielding requires a denser cloud and a correspondingly higher evaporation rate. Furthermore, it was assumed that the lifetimes of the ions in circular orbits are limited by Poynting-Robertson drag only. In reality, since  the effective gravitational constant $G(1-f)$ depends on the particular ionic state, different ions on a circular orbit with given radius can collide inelastically. This reduces the lifetimes and  further increases  the  evaporation rate. Necessarily, the particles of the gas also collide inelastically with the  fast particles after their evaporation. Since this may increase the energy of particles of the cloud, the sign of this influence on the expected evaporation rate is not trivial. Note that $\nu m$ is a time average of a mass evaporation rate, which is strongly peaked near the perihel of Z.

\section{Mean free path of the evaporated particles}

In order to explain that the evaporated particles do not reach the distance of Earth's orbit, we supposed that the particles first rapidly loose their energy by inelastic collisions and  then  slowly by Poynting--Robertson drag. From this a minimum evaporation rate follows. These assumptions become self-consistent, provided the mean free path for collisions within the cloud is small compared to its radius $a_{final}$. The radial size of an atom is about Bohr's radius $ r_B=0.53\cdot 10^{-10}$ m, so the scattering crossection for two atoms becomes $\Sigma = \pi\cdot (2 r_B)^2=3.5\cdot 10^{-20}$ m$^2$. We shall use this number for inelastic scattering, although the scattering of ions is more involved. Between the distances $r+dr$ and $r$ the number of particles is $\nu \beta 2\pi r dr$, and the volume  $4\pi r^2 dr \sin (\alpha )$, so that the particle density $D$ at distance $r$ becomes
\begin{equation}
D = {\nu \beta\over 2 r\sin (\alpha )}.
\end{equation}
Numerically, for $r=a_{final}=8\cdot 10^9$ m, i.e. at the outer edge of the cloud, $D={1\over f}\cdot 4.5\cdot 10^{10}$ m$^{-3}$, which depends on the (average) reciprocal value of $f$. The mean free path for collisions for an evaporated particle is of the order of 
\begin{equation}
\lambda = {1\over \Sigma D}= {f\cdot 6\cdot 10^{8}}\mbox{ m},
\end{equation}
where $f$ belongs to particles of the cloud.
Since for bound particles $f<1$ and for ions $f\ll 1$, we have $\lambda \ll a_{final}$. This garantees the self-consistency of the assumed rapid energy loss by collisions. 

It may be a more delicate question, whether some ions  will not capture electrons and become atoms of a species with$f>1$ or small molecules, for which always $f>1$, so that these can escape. It cannot be stressed enough that the dynamics of the gas cloud  may turn out to be a very complex theoretical problem. It is the basis for understanding the temperature changes of the ice age era. 

\section{The time dependent cloud}

The small cloud around the Sun, in which the emitted particles rapidly loose their initial high energy by collisions, was found to be self-consistent. The question remains, whether this is what establishes itself. Starting without a cloud, the evaporated particles are not stopped by collisions. If they are ions, any electric or magnetic field could deviate them from their path, which is $f$-dependent. The cloud may be spread over rather vast space so that for a time the particles do not collide.  However, since the influx of particles continues, their density increases. At a certain point, energy loss by inelastic collisions becomes appreciable. Therefore, the volume of the cloud shrinks and the collisions become more frequent. A  transition from  individual motion to collision-dominated energy loss must necessarily occur in a time that is shorter than the Poynting-Robertson lifetime. As indicated by Eq.~[\ref{lifetime}] (valid for circular  motion) this lifetime increases with the square of the size of the orbit. For an estimate,  using the mass of an O-atom, and $a= 1.5\cdot 10^{11}$ m (Earth's distance from the Sun), we get $t= {400\over f}$  years, where we expect $f\ll 1$  for ions.

As the particles  move closer to the Sun, the amount of light scattered per particle increases. Inelastic scattering favours the formation of a disk. If Earth's orbit is near the midplane of the cloud, which is presumably close to the invariant plane of the planetary system, the insolation is diminished \cite{Muller1,Muller2}.  In the endphase of the cloud, the particles have aligned motions in small orbits, and they spiral into the Sun due to Poynting-Robertson drag. Thus the initial dilute cloud gradually becomes denser until collisions induce a collaps. The slowly growing shielding is followed by a rather sudden return to the full solar insolation. This, of course,  will be followed by another buildup of a cloud. Such a time dependence of the shielding reproduces characteristic features of the Dansgaard-Oeschger temperature peaks. 21 such events were counted  between 90 and 11.5 kyr BP \cite{GRIP,North}, which corresponds to 4000 yr as the average spacing between two events. More detailed data reveal a period of 1470 yr \cite{Rahmstorf}. The interval is with both values too short to be connected with Milankovitch modifications of orbits. On the other hand, if the particles of the time dependent cloud are ions, then their values $f$ are expected to be sufficiently small, so that the Poynting-Robertson lifetime, equation [\ref{lifetime}], is longer than this period. Therefore,  the period may be determined by the dynamics of the cloud. Tentatively, we explain the Dansgaard-Oeschger temperature variations with  sequences of formations and collapses of the screening cloud. 

Extensive studies will be necessery to establish the conditions under which either the small cloud or the time dependent cloud is real. We are aware that we left aside plasma properties, electric and magnetic fields and the question, whether energy loss via gas discharges might play a role. The size of the small cloud was determined by the high excentricity of the initial orbits of evaporated particles. The smallest semiaxis compatible with the angular momentum defines the size of the small cloud. We are not aware of a further physical lenght that could prevent the time dependent cloud from reaching the Earth. Although the cloud is dilute initially, a particle flux into Earth's atmosphere seems unavoidable. It would be important to establish upper bounds for extraterrestrial components in ice cores. The evaporation from Z is limited by the escape velocity and it acts as a mass-sensitive fractional distillation. Therefore, a measurement of the isotopic ratio of a stable impurity element in an ice core could define such a bound.     

\section{Unavoidable ice age era}

In this model, Z is necessary for the polar shift, but Z does not exist any more. This is indeed a strongly limiting condition for the scenario. Clearly, Z had to be hot and emitting material already before the polar shift. This period lasted a few million cycles, the estimated number of trials for a  narrow approach  to Earth to occur. During this time Earth's climate was influenced by the screening of the solar radiation by the  gas cloud. The existence of a polar shift at the end of the Pleistocene, connected with the condition, that the culprit has disappeared, leads to the prediction of a variable cold period lasting a few million years. The  Milanchovitch theory without polar shift does not predict this. It rather predicts variations in insolation during an unlimited time forward and backward. It cannot cope with the observed fact that before about 3 million years BP the climate was warm and had only small fluctuations \cite{Tiedemann}. The global temperature was similar to that of the Holocene. 

\section{Cloud after the pole shift}

Since the evaporation rate increased enormously after the narrow encounter, what happened with the cloud? The pole shift occured towards the end of the Pleistocene, probably before the Younger Dryas, which was a cold period, but not distinctly colder than previous periods. Now, the high velocities of Earth and Z were determined by their motion in the Sun's large gravitational field.  If the relative velocity was of the order of 40 km/s, then the momentum transfer during the narrow encounter produced a modest scattering of Z  by an angle of a few degrees. Before, the orbital planes of Earth and Z probably almost coincided. After the scattering, the orbital plane of Z likely was rotated by a few degrees around an axis that passed through the location of the close encounter and the Sun. This tilt saved the Earth from an extreme shielding of the solar radiation. The evaporated material (mostly in molecular form) moved outside the existing disk of ionic particles. Thus it was blown out of the planetary system by radiation pressure. Even if a new ionic cloud established itself in the tilted plane, this shielding affected the Earth only twice a year, i.e. when the Earth crossed the plane of the cloud. Again, more detailed modelling might establish probabilities of various occurences. It seems likely that the scattering of the fractions of Z during the narrow encounter saved the Earth from dramatic shielding and influx of material.

\section{Conclusions}

Remains of mammoths in arctic Siberia, where there is not enough sunlight per year to grow the plants that feed these animals, prove, that the latitude of this region  was lower in the Pleistocene. Therefore a scenario must exist, which leads to this geographic polar shift. The paper describes a proposal which inolves an additional planet. In a close encounter between this planet and the Earth the tidal force created a 1 permil extensional deformation on the Earth. While this deformation relaxed to a new equilibrium shape in a time of at least 200 days, the precessional motion of the globe resulted in the geographic polar shift. Since the additional planet does not exist any more, it must have moved in an extremely eccentric orbit, so that it was hot and evaporating. It produced a disk-shaped gas cloud around the Sun which partially shielded the solar radiation. The resulting glaciation on Earth depended on the inclination of Earth's orbit relative to the invariant plane, which has a period of about 100 kyr. We describe two cloud structures, namely a time independent cloud and a cloud, whose density increases until inelastic collisions induce its collaps. Tentatively we assume that the resulting time dependence of the screening  produced the Dansgaard-Oeschger temperature peaks. The Heinrich events might result from earthquakes accompaniing passages between Earth and planet up to about Moon's distance.  During the  polar shift event the Earth must have torn the other planet to peaces which subsequently evaporated completely. Such a  close encounter is expected to occur only once in several millions of orbital motions of the planet. Therefore the model correctly predicts a duration  for the era of glaciations of the order of the length of the Pleistocene. We conclude that it is possible to construct a scenario which leads to the required geographic polar shift. The resulting model has the potential to describe the dominant observed features of the Pleistocene glaciations. It involves complex problems  which here are treated only rudimentarily. More elaborate studies would be very valuable. In the case that these contradict the given estimates, they could only question this particular scenario, but not the conclusion that a pole shift has occured.

\section*{Acknowledgement}
We wish to express our gratitude to Hans-Ude Nissen, who has been a critical and knowledgeable discussion partner throuout this research.

\end{document}